\documentclass[aps,prl,twocolumn,showpacs,superscriptaddress]{revtex4}
\usepackage{graphicx}

\begin{document}
\title{Experimental Demonstration of a Quantum Protocol\\
for Byzantine Agreement and Liar Detection}
\author{Sascha Gaertner}
\email{ssg@mpq.mpg.de}
\affiliation{Max-Planck-Institut f\"{u}r Quantenoptik, D-85748 Garching, Germany}
\affiliation{Fakult\"{a}t f\"{u}r Physik, Ludwig-Maximilians-Universit\"{a}t,
D-80799 M\"{u}nchen, Germany}
\author{Mohamed Bourennane}
\affiliation{Department of Physics, Stockholm University,
SE-10691 Stockholm, Sweden}
\author{Christian Kurtsiefer}
\affiliation{Department of Physics, National University of Singapore,
117542 Singapore, Singapore}
\author{Ad\'an Cabello}
\email{adan@us.es}
\affiliation{Departamento de F\'{\i}sica Aplicada II, Universidad de Sevilla,
E-41012 Sevilla, Spain}
\author{Harald Weinfurter}
\affiliation{Max-Planck-Institut f\"{u}r Quantenoptik, D-85748 Garching, Germany}
\affiliation{Fakult\"{a}t f\"{u}r Physik, Ludwig-Maximilians-Universit\"{a}t,
D-80799 M\"{u}nchen, Germany}




\begin{abstract}
We introduce a new quantum protocol for solving detectable Byzantine
agreement (also called detectable broadcast) between three parties,
and also for solving the detectable liar detection problem. The
protocol is suggested by the properties of a four-qubit entangled
state, and the classical part of the protocol is simpler than that
of previous proposals. In addition, we present an experimental
implementation of the protocol using four-photon entanglement.
\end{abstract}



\maketitle


A basic goal in distributed computing is to achieve coordination
despite the failure of some of the distributed processes. This
requires the nonfaulty components to reach an agreement. The
problem of coping with such tasks is expressed abstractly as the
Byzantine Generals Problem, also called Byzantine Agreement (BA)
\cite{PSL80,LSP82}.

Three divisions of the Byzantine army, each commanded by its own
general, are besieging an enemy city. The three generals $A$, $B$,
and $C$ can communicate with one another by messengers only (i.e., by
pairwise authenticated error-free classical channels). They must
decide upon a common plan of action either $0$ or $1$ (for instance,
attack or retreat). The commanding general $A$ decides on a plan and
communicates this plan to the other two generals by sending $B$ a
message $m_{AB}$ (either $0$ or $1$), and by sending $C$ a message
$m_{AC}$. Then, $B$ communicates the plan to $C$ by sending him a
message $m_{BC}$, and $C$ communicates the plan to $B$ by sending
him a message $m_{CB}$. However, one of the generals (including $A$)
might be a traitor, trying to keep the loyal generals from agreeing
on a plan. The BA problem is to devise a protocol in which (i) all
loyal generals follow the same plan, and (ii) if $A$ is loyal, then
every loyal general follows the plan decided by $A$. From the point
of view of a loyal $C$ receiving different messages from $A$ and
$B$, the BA problem is equivalent to the liar detection problem
\cite{Cabello02}, in which $C$'s task is to ascertain who is lying,
$A$ or $B$.

The BA problem has been proven to be unsolvable \cite{PSL80,LSP82},
unless each of the generals is in possession of a list of numbers
unknown to the other generals, but suitably correlated with the
lists of the other generals. Therefore, solving the BA problem can
be reduced to solving the problem of the generation and secure
distribution of these lists. A quantum protocol enables one to test
the security of the distribution, however, in case of an attack, no
secret lists are available and the whole communication has to be
aborted. Still, in this case, a variation of the BA, called
detectable Byzantine agreement (DBA) or detectable broadcast
\cite{FGM01} can be solved \cite{FGM01}. In the DBA problem,
conditions (i) and (ii) are relaxed so that (i') either all loyal
generals perform the same action or all abort, and (ii') if $A$ is
loyal, then either every loyal general obeys the order sent by $A$
or aborts. Consequently, we can define a protocol for solving the
detectable liar detection problem as that one in which the possible
outcomes for a loyal $C$ receiving different messages from $A$ and
$B$ are either to detect who is lying or to abort
\cite{Cabello02,Cabello03a,Cabello03b}.

The properties of two specific entangled states have suggested two
different methods for solving the DBA problem. The first method was
inspired by the properties of the three-qutrit singlet state, and it
is based on lists of six combinations of numbers \cite{FGM01}. Such
lists can also be distributed using two quantum key distribution
protocols \cite{IG05}. The second method was suggested by the
properties of a four-qubit entangled state, and it is based on lists
of four combinations of numbers \cite{Cabello03b}.

In this Letter we introduce a new protocol for solving the DBA
problem. It uses simpler lists than those in \cite{FGM01,IG05}, and
uses them more efficiently than in \cite{Cabello03b}. In contrast to
\cite{IG05}, it allows the simultaneous generation of all lists. In
addition, we present the first experimental demonstration of a quantum
protocol for DBA and liar detection via four-photon entanglement.



\begin{figure}
\centerline{\includegraphics[width=0.74\columnwidth]{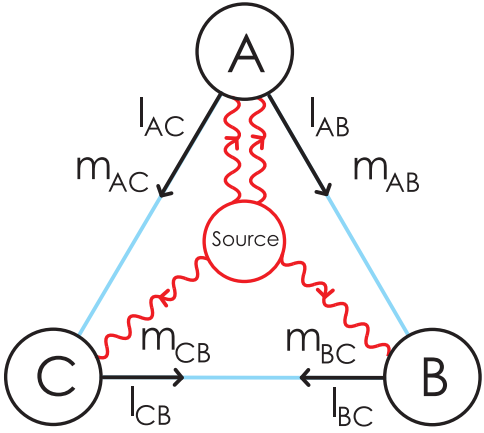}}
\caption{Quantum protocol for detectable Byzantine agreement. Three
generals, $A$ (the commanding general), $B$, and $C$, are connected
by pairwise authenticated error-free classical channels. In the
first part of the protocol, four qubits prepared in the state
$|\Psi^{(4)}\rangle$ are distributed among the parties and, after a
classical discussion, either (a) each general obtains a list $l_i$,
or (b) all loyal generals agree to abort. If (a) then, in the second
part of the protocol, $A$ sends $B$ ($C$) a message $m_{AB}$
($m_{AC}$) and a list $l_{AB}$ ($l_{AC}$), and $B$ ($C$) sends $C$
($B$) a message $m_{BC}$ ($m_{CB}$) and a list $l_{BC}$ ($l_{CB}$).}
\label{scheme}
\end{figure}

The protocol has two parts. The goal of the first part is to
generate and distribute three lists, $l_{A}$ for $A$, $l_{B}$ for
$B$, and $l_{C}$ for $C$ utilizing the characteristic properties of
a particular four-photon polarization entangled state
\cite{WZ01,EGBKZW03,GBEKW03}, and to check for the security of this
distribution. Once the parties have these lists, in the second part
of the protocol they use them, together with pairwise classical
communication, for reaching the agreement (Fig.~\ref{scheme}). The
option to abort will be used only in the distribution part.
Thereafter, the protocol enables full BA.


In detail, the lists $l_A$, $l_B$, and $l_C$ have the following
properties \cite{Cabello03b}:
(I) The three lists have the same length $L$. The elements of
$l_{A}$ are random trits (i.e., $0$, $1$, or $2$). The elements of
$l_{B}$ and $l_{C}$ are random bits (i.e., $0$ or $1$).
(II) At position $j$ in these lists, we find the combinations $000$
(i.e., ${l_A}_j=0$, ${l_B}_j=0$, ${l_C}_j=0$), $111$, or, with equal
probability, either $201$ or $210$.
(III) Each party cannot know other parties' lists beyond what can be
inferred from his own list and properties (I) and (II).

The result of this first part can be that (a) all parties agree that
they have the right lists and can start the second part of the
protocol or (b) agree to abort it.

To simplify the discussion of the second part of the protocol, note
that the roles of $B$ and $C$ are symmetrical, and thus everything
we say about $B$ applies to $C$ and vice versa. The agreement part
runs as follows:
(i) When $A$ sends $m_{AB}$, this message must be accompanied by
other data which must be correlated with the message itself and, at
the same time, must be known only by $A$. For that purpose, $A$ also
sends $B$ a list $l_{AB}$ with all the positions in $l_{A}$ in which
the value $m_{AB}$ appears. After that, if $A$ is loyal he will
follow his own plan.

Example: if $A$ is loyal, the message is $m_{AB}=m_{AC}=0$, and
$A$'s list is $l_A=\{2, 0, 0, 2, 1, 1, 0, 0, 2, \ldots\}$, then $A$
must also send $l_{AB}=l_{AC}=\{{\bf 2}, {\bf 3}, {\bf 7}, {\bf 8},
\ldots\}$.

When $B$ receives $m_{AB}$ and $l_{AB}$, only one of two things can
happen:
(ia) If $l_{AB}$ is of the appropriate length [i.e., approximately
$L/3$, according to property (I)], and $m_{AB}$, $l_{AB}$, and
$l_{B}$ do satisfy (II), then we will say that the data (i.e.,
$m_{AB}$, $l_{AB}$, and $l_{B}$) are {\em consistent}. If the data
are consistent, then $B$ will follow the plan $m_{AB}$ unless $C$
convinces him that $A$ is the traitor in the next step of the
protocol [see (ii)].
(ib) If $m_{AB}$, $l_{AB}$, and $l_{B}$ are inconsistent, then $B$
ascertains that $A$ is the traitor, and $B$ will not follow any plan
until he reaches an agreement with $C$ in the next step of the
protocol [see (ii)].

Example: $B$ would receive inconsistent data if he receives the
message $m_{AB} = 0$ accompanied by the list $l_{AB}=\{{\bf 2}, {\bf
5}, {\bf 6}, {\bf 7}, \ldots\}$, and $B$'s list is $l_{B}=\{1, 0, 0,
0, 1, 1, 0, 0, 0, \ldots\}$. This data is inconsistent because
$l_{A}$ cannot have $0$ at positions ${\bf 5}$ and ${\bf 6}$.

(ii) The message $m_{BC}$ can be not only $0$ or $1$, but also
$\bot$, meaning ``I have received inconsistent data.'' If the
message is $0$ or $1$, it must be accompanied by other data which
prove that $m_{BC}$ is the same one that $B$ has received from $A$;
i.e., data that $B$ could only have obtained from $A$ if
$m_{BC}=m_{AB}$. For that purpose, $B$ also sends $C$ a list
$l_{BC}$ which is supposedly the same list $l_{AB}$ that $B$ has
received from $A$.

When $C$ receives $m_{BC}$ and $l_{BC}$, he already has $m_{AC}$ and
$l_{AC}$. Then, only one of six things can happen:
(iia) If $m_{AC}$, $l_{AC}$, and $l_{C}$ are consistent, and
$m_{BC}$, $l_{BC}$, and $l_{C}$ are also consistent, and
$m_{AC}=m_{BC}$, then $C$ will follow the plan $m_{AC}=m_{BC}$.
(iib) If $m_{AC}$, $l_{AC}$, and $l_{C}$ are consistent, and
$m_{BC}$, $l_{BC}$, and $l_{C}$ are also consistent, but $C$ is
receiving conflicting messages ($0$ or $1$) from $A$ and $B$, then
$C$ ascertains that $A$ is the traitor and $B$ is loyal, since $A$
is the only one who can send consistent data to $B$ and $C$. Since
the roles of $B$ and $C$ are symmetrical, $B$ also ascertains that
$A$ is the traitor and $C$ is loyal. Then $C$ and $B$ will follow a
previously decided plan, for instance, $0$.
(iic) If $m_{AC}$, $l_{AC}$, and $l_{C}$ are consistent, and $C$ is
receiving $m_{BC}=\bot$, then $C$ will follow the plan $m_{AC}$.
Note that in this case there is no way for $B$ to convince $C$ that
he has actually received inconsistent information from $A$.
Therefore, following the plan $m_{AC}$ (even if $A$ is the traitor)
is the only option for reaching agreement with the other loyal
party.
(iid) If $m_{AC}$, $l_{AC}$, and $l_{C}$ are consistent, but
$m_{BC}$, $l_{BC}$, and $l_{C}$ are inconsistent, then $C$
ascertains that $B$ is the traitor and $A$ is loyal. Then $C$ will
follow the plan $m_{AC}$.
(iie) If $m_{AC}$, $l_{AC}$, and $l_{C}$ are inconsistent, but
$m_{BC}$, $l_{BC}$, and $l_{C}$ are consistent, then $A$ is the
traitor. Then, complementary to case (iic), they will now follow the
plan $m_{BC}$.
(iif) If $m_{AC}$, $l_{AC}$, and $l_{C}$ are inconsistent, and $C$
is receiving $m_{BC}=\bot$, this means that both $C$ and $B$ know
that $A$ is the traitor. Then $C$ and $B$ will follow the previously
decided plan $0$.




The generation and distribution of the lists with properties (I),
(II), and (III) is achieved by distributing among the parties four
qubits initially prepared in some specific state, then making local
single qubit measurements on the four qubits, and then testing
(using the pairwise classical channels) whether or not the results
of these measurements exhibit the required correlations.

The state used in our protocol is the four-qubit state
\begin{eqnarray}
|\Psi^{(4)}\rangle_{abcd} & = & {\frac{1}{2 \sqrt{3}}}
(2|0011\rangle-|0101\rangle-|0110\rangle-|1001\rangle \nonumber \\
& & -|1010\rangle+2|1100\rangle)_{abcd},
\end{eqnarray}
where, e.g., $|0011\rangle_{abcd}$ means $|0\rangle_a \otimes
|0\rangle_b \otimes |1\rangle_c \otimes |1\rangle_d$. This state has
been observed in recent experiments \cite{GBEKW03,BEGKCW04}. The
protocol exploits two properties of this state, i.e., the fact that
it is invariant under the same unitary transformation applied to the
four qubits (i.e., $U \otimes U \otimes U \otimes U
|\Psi^{(4)}\rangle_{abcd} = |\Psi^{(4)}\rangle_{abcd}$), where $U$
is any unitary operation acting on one qubit, and the fact that it
exhibits the required perfect correlations between the results of
projection measurements on the four qubits. Specifically, if $A$
measures qubits (a) and (b), $B$ measures qubit (c) and $C$ measures
qubit (d), and all of them are measuring in the same basis, then: if
the results of the measurements on qubits (a) and (b) are both $1$
(which $A$ will record as a single $0$)
---something which occurs with probability \mbox{$1/3$---,}
then the result of the measurement on qubit (c) must be $0$ (which
$B$ will record as $0$) and the result of the measurement on qubit
(d) must be $0$ (which $C$ will record as $0$). If the results of
the measurements on qubits (a) and (b) are both $0$ (which $A$ will
record as a single $1$), then the result of the measurement on qubit
(c) must be $1$ (which $B$ will record as $1$) and the result of the
measurement on qubit (d) must be $1$ (which $C$ will record as $1$).
Finally, if the results of the measurements on qubits (a) and (b)
are either $0$ and $1$, or $1$ and $0$ (which $A$ will record as a
single $2$), then the results of the measurements on qubits (c) and
(d) can be either $0$ and $1$, or $1$ and $0$.

The distribute and test part of the protocol consists of the
following steps:
{\em (i)} A source emits a large number of four-qubit systems in the
state $|\Psi^{(4)}\rangle$. For each four-qubit system $j$, qubits
(a) and (b) are sent to $A$, qubit (c) to $B$ and qubit (d) to $C$.
{\em (ii)} For each four-qubit system $j$, each of the three parties
randomly chooses between two projection measurements; e.g., each of
them either measures in the $\{|0\rangle,|1\rangle\}$ basis or in
the $\{|\bar{0}\rangle, |\bar{1}\rangle\}$ basis [where
$|\bar{0\rangle} = (|0\rangle + |1\rangle)/\sqrt{2}$ and
$|\bar{1}\rangle = (|0\rangle - |1\rangle)/\sqrt{2}$] and makes a
list with his results. To extract the correlated fourfold
coincidences, they do the following. For the first one third of the
experiments, $C$ asks $A$ and $B$ whenever they have detected and in
which bases they have measured their qubits ($50\%$ of the cases,
$A$ speaks first, and in the other $50\%$, it is $B$ who speaks
first). Then, $C$ tells $A$ and $B$ which events should be rejected.
For the second one third of the experiments, $B$ and $C$ exchange
their roles, and for the last one third, $A$ and $B$ exchange their
roles. By exchanging the roles, they ensure that none of the generals
can fake parts of the classical protocol without being discovered.
After this step, each of the parties has a list. These lists are all
of the same length. $A$ has a list $l_{A}$ of trits, and each of $B$
and $C$ has a list, $l_{B}$ and $l_{C}$ respectively, of bits.
{\em (iii)} $C$ randomly chooses a position $k_{C}$ from his list
$l_{C}$ and asks $A$ and $B$ to inform him, via the pairwise
classical channels, about their results on the same position
$k_{C}$. If all parties have measured in the same basis, their
results must be suitably correlated. After this step, each party
discards the entries in their lists which were used for this test.
{\em (iv)} The parties exchange their roles; i.e., $B$ randomly
chooses a new position $k_{B}$ from his list and repeats step {\em
(iii)}; then $A$ chooses a new position $k_{A}$, etc. $C$ starts the
process all over again until a large number of tests have been
performed.

This part of the protocol has only two possible outcomes: Depending
on the observed quantum error ratio (QER), defined as the ratio of
incorrect/all four-photon detection events, the loyal generals
decide to abort or use the lists $l_A$, $l_B$, and $l_C$ to reach
the agreement.





\begin{figure}
\centerline{\includegraphics[width=0.92\columnwidth]{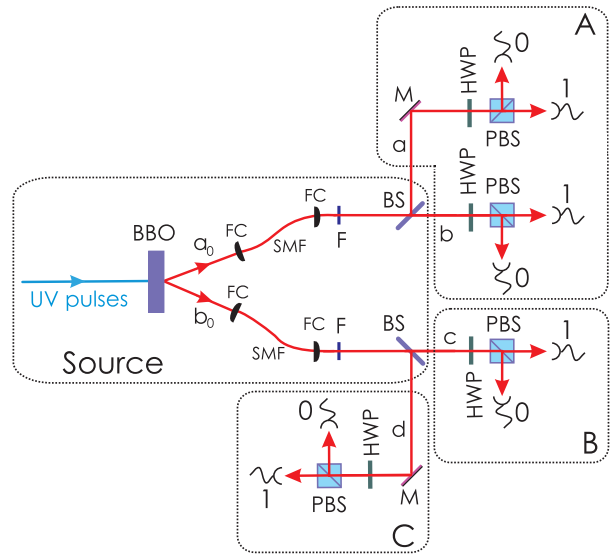}}
\caption{Scheme of the experimental setup. UV-pulses pump a
beta-barium borate crystal $BBO$. The degenerate down-conversion
emission into the two directions, $a_0$ and $b_0$, is coupled into
optical fibers by fiber couplers $FC$, then passes interference
filters $F$. To generate the state $|\Psi^{(4)}\rangle$, the initial
emission modes are split with two nonpolarizing beam splitters
$BS$. Two of the photons are sent to $A$, one to $B$, and one to
$C$. Then, each party performs polarization measurements by
inserting a half-wave plate $HWP$ and using a polarizing beam
splitter $PBS$ and single-photon avalanche detectors.} \label{setup}
\end{figure}


In the experimental implementation, the physical qubits are polarized
photons, and the states $|0\rangle$ and $|1\rangle$, correspond,
respectively, to the vertical and horizontal linear polarization
states, $|V\rangle$ and $|H\rangle$. To prepare the state
$|\Psi^{(4)}\rangle$, we have used the emission of four photons
produced in the second order of perturbation of the type-II process
of spontaneous parametric down-conversion
\cite{WZ01,EGBKZW03,GBEKW03}. The experimental setup is shown in
Fig.~\ref{setup}. We have used UV-pulses of a frequency doubled
mode-locked Titan:Sapphire laser (pulse length $130$~fs and
repetition rate $82$~MHz) to pump a $2$~mm thick beta-barium borate
(BBO) crystal at a wavelength of $390$~nm and with an average power
of $750$~mW. The pump beam has been focused to a waist of 100~$\mu$m
inside the crystal. The degenerate down-conversion emission into the
two characteristic type-II crossing directions, $a_0$ and $b_0$, has
been coupled into single mode optical fibers (length $2$~m) to
precisely define the spatial emission modes. After the fibers, the
down-conversion light has passed interference filters with a
bandwidth of $3$~nm. To generate the four-photon state
$|\Psi^{(4)}\rangle$, the initial emission modes have been split
with two nonpolarizing beam splitters. We have selected those
events in which one photon is detected in each of the resulting four
outputs ($a$, $b$, $c$, and $d$) of the beam splitters.

The polarization measurements have been performed by inserting
half-wave plates in each of the four modes. For measuring in the
polarization bases $\{|H\rangle,|V\rangle\}$ and
$\{(|H\rangle+|V\rangle)/\sqrt{2},(|H\rangle-|V\rangle)/\sqrt{2}\}$,
the orientations of the half-wave plates have been randomly switched
between $0^\circ$ and $22.5^\circ$ respectively. The switching of
the wave plates has been controlled by random number generators. The
registration time for a fixed setting has been $1$~s. The four
photons have been detected, after passing polarizing beam splitters,
by eight passively quenched single-photon Si-avalanche photodiodes
and registered with an eight-channel multiphoton coincidence
counter, which allows an efficient registration of the $16$~relevant
fourfold coincidences \cite{GKW05}. When more than one four-photon
coincidence has been recorded in the same time window, only the
first one has been used. To translate the detection events into bit
values, we have associated a single-photon detection in the
reflected (transmitted) output port of the polarization beam
splitters with the bit value $0$ ($1$). All the detection events and
the basis settings have been registered with a personal computer.




To generate the lists, the parties have performed $48184$
measurements in $17$ hours. To extract the fourfold coincidences in
each time window, each party has asked the other parties whenever
they detected a photon. After removing those entries where they have
not registered a photon, they have obtained lists $l_{A}$, $l_{B}$
and $l_{C}$ with $12043$ entries containing $3000$ correlated
bits/trits with a QER of $5.47\%$. For the first part of the
protocol, each of the parties has randomly chosen $1000$ entries
from his list. To check whether their results are perfectly
correlated or not, each party has computed the QER for those entries
which should be perfectly correlated from his subset. $A$ has
obtained a QER of $3.32\%$, $B$ $4.64\%$, and $C$ $5.40\%$ (the QERs
depend on the randomly chosen subsets). For the second part of the
protocol, the parties have used the remaining correlated entries of
their lists. A subset of these lists is shown in Table~\ref{tab}.

In conclusion, we have introduced a new quantum protocol for solving
a fundamental problem in fault-tolerant distributed computation and
database replication. Our protocol uses simpler lists or uses them
more efficiently than previous protocols, and permits the
simultaneous generation of all the lists. In addition, we have
presented the first experimental demonstration of a quantum protocol
for DBA and liar detection via four-qubit entanglement. Although the
same problems could be solved by linking several quantum key
distribution protocols, our results show that a more specific and
elegant quantum solution requiring a subtler form of entanglement is
feasible with present technology.




\begin{acknowledgments}
The authors thank N. Gisin and M. \.{Z}ukowski for useful
conversations. This work was supported by DFG, the Swedish Research
Council (VR), the Spanish MEC Project No. FIS2005-07689, and the EU 6FP
program QAP.
\end{acknowledgments}


\begin{table}[t]
\caption{\label{tab} Part of the lists $l_{A}$, $l_{B}$, and $l_{C}$
obtained experimentally. Numbers in italics are events which should
not occur in an ideal case.}
\begin{ruledtabular}
{\begin{tabular}{cccccccc}
${\rm Position}$ & $l_A$ & $l_B$ & $l_C$ & ${\rm Position}$ & $l_A$ & $l_B$ & $l_C$ \\
\hline
${\bf 1}$ & $2$ & $1$ & $0$ & ${\bf 16}$ & $1$ & $1$ & $1$ \\
${\bf 2}$ & $0$ & $0$ & $0$ & ${\bf 17}$ & $1$ & $1$ & $1$ \\
${\bf 3}$ & $0$ & $0$ & $0$ & ${\bf 18}$ & $1$ & $1$ & $1$ \\
${\bf 4}$ & $2$ & $0$ & $1$ & ${\bf 19}$ & $1$ & $1$ & $1$ \\
${\bf 5}$ & $1$ & $1$ & $1$ & ${\bf 20}$ & $0$ & $0$ & $0$ \\
${\bf 6}$ & $1$ & $1$ & $1$ & ${\bf 21}$ & $2$ & $1$ & $0$ \\
${\bf 7}$ & $0$ & $0$ & $0$ & ${\bf 22}$ & $0$ & $0$ & $0$ \\
${\bf 8}$ & $0$ & $0$ & $0$ & ${\bf 23}$ & $2$ & $0$ & $1$ \\
${\bf 9}$ & $2$ & $0$ & $1$ & ${\bf 24}$ & $0$ & $0$ & $0$ \\
${\bf 10}$ & $2$ & $0$ & $1$ & ${\bf 25}$ & $2$ & $1$ & $0$ \\
${\bf 11}$ & $2$ & $1$ & $0$ & ${\bf 26}$ & $1$ & $1$ & $1$ \\
${\bf 12}$ & $2$ & $0$ & $1$ & ${\bf 27}$ & $1$ & $1$ & $0$ \\
${\bf 13}$ & $0$ & $0$ & $0$ & ${\bf 28}$ & ${\it1}$ & ${\it1}$ & ${\it0}$ \\
${\bf 14}$ & ${\it2}$ & ${\it1}$ & ${\it1}$ & ${\bf 29}$ & $2$ & $1$ & $1$ \\
${\bf 15}$ & $2$ & $0$ & $1$ & ${\bf 30}$ & $2$ & $0$ & $1$ \\
\end{tabular}}
\end{ruledtabular}
\end{table}



\end{document}